\documentclass{article}
\usepackage[utf8]{inputenc}

\usepackage{graphicx,subfigure}
\usepackage{amsmath}
\usepackage{longtable,tabularx}
\usepackage{amssymb,amsfonts}

\usepackage{amssymb,dsfont}
\usepackage{amsmath,color}%
\usepackage{amsfonts}%
\usepackage{graphicx,subfigure}
\usepackage{authblk} 
\providecommand{\U}[1]{\protect\rule{.1in}{.1in}}

\usepackage{algorithm}
\usepackage{algorithmicx}
\usepackage[noend]{algpseudocode}

\newcommand{\beq} {\begin{eqnarray*}}
\newcommand{\eeq} {\end{eqnarray*}}

\title{Prediction of Preliminary Maximum Wing Bending Moments under Discrete Gust}
\date{}
\author[1,2,4]{Edouard Fournier\thanks{ \texttt{edouard.fournier@airbus.com } }}
\author[2]{  St\'ephane Grihon\thanks{ \texttt{stephane.grihon@airbus.com } }} 
\author[3]{Christian Bes \thanks{ \texttt{christian.bes@univ-tlse3.fr  } }}
\author[1,4]{Thierry Klein\thanks{ \texttt{thierry.klein@math.univ-toulouse.fr  or thierry01.klein@enac.fr} } } 
\affil[1]{Institut de math\'ematique, UMR5219; Universit\'e de Toulouse;  CNRS, UPS IMT, F-31062 Toulouse Cedex 9, France} 
\affil[2]{Airbus France 316, Route de Bayonne, Toulouse France} 
\affil[3]{Univerit\'{e} de Toulouse, Mechanical Engineering Department, 118 Route de Narbonne, Cedex 4 Toulouse} 
\affil[4]{ENAC - Ecole Nationale de l'Aviation Civile, Universit\'e de Toulouse, France}

\begin{document}

\maketitle

\section{Introduction}

Many methodologies (see \cite{CB1,CB2,CB3,CB4,CB5,CB6,CB7,papila2000}), have been proposed to quickly identify among a very large number of flight conditions and maneuvers (i.e., steady, quasi-steady and unsteady loads cases) the ones which give the worst values for structural sizing (e.g., bending moments, shear forces, torques,...). All of these methods use both the simulation model of the aircraft under development and efficient algorithms to find out the critical points of the flight envelope. At the preliminary structural design phases detailed models are not available and airframe's loads are estimated by empirical relationships or engineering judgments. These approximations can induce load uncertainties and may lead to expensive redesign activities through the upcoming detailed sizing process (see \cite{lomax1996,howe2004}). In the context of preliminary design phase for a weight aircraft variant without geometric change, to overcome this likely drawback, we propose a method based on the huge and reliable database of an initial aircraft from which the weight variant belongs. More precisely, from the load cases of this initial database, response surfaces are identified as functions of preliminary parameters (flight conditions and structural parameters). Then, these response surfaces are used to predict quickly the weight aircraft variant quantities of interest for preliminary structural design studies. Although the proposed method can be readily extended to any structural quantity of interest and to any flight conditions and maneuvers, it is presented here for the prediction of the bending moments due to discrete gust at different locations along a wing span. \\

This note is organized as follows. Section 2 presents the initial aircraft database where its values are derived from a detailed aeroelastic model. This database is composed of the maximum temporal value of the bending moment due to the discrete gust at any wing span location and for any point inside the flight envelope. These maximum values are identified by few preliminary parameters (altitude, mass, speed, gust length, etc.). Section 3, describes the Orthogonal Greedy Algorithm (OGA) which permits to obtain the coefficients of the response surfaces from the initial database as a function of the preliminary parameters. This algorithm is based on parsimony principle and aims to protect against under and over fitting of the response surface when it is used to predict the maximum bending moment for a weight variant aircraft. Section 4 presents results of the predictions and the confidence bounds of the expected maximum temporal bending moment along the wing span for weight aircraft variants. These surface response predictions are further compared and validated with existing weight aircraft variants database results. Finally, in Section 5, conclusion and future works are presented.

\section{Initial Aircraft database}

The database of the initial aircraft wing at hand is built from a detailed aeroelastic model which have been updated from ground and flight tests. This aeroelastic model under gust  \cite{CB13,CB11,CB4} can be expressed in the Laplace domain as follows

\begin{equation}
(s^{2}\textbf{M} + s\textbf{C}+ K-q_{\infty}\textbf{Q}_{GG}(s))z(s)=\frac{q_{\infty}}{\textbf{V}}\textbf{Q}_{Gg}(s)u(s)
\label{eq:eq00}
\end{equation}

with $u(s)$ is the gust sequence, $z(s)$ is the structural response, $\textbf{K}$, $\textbf{C}$, $\textbf{M}$ are respectively the stiffness, the damping and the mass matrices, $\textbf{Q}_{GG}$ is the motion-induced unsteady aerodynamic matrix, $\textbf{Q}_{Gg}$ is the gust-induced unsteady aerodynamic matrix, $q_{\infty}$ is the freestream dynamic pressure, $V$ is the the air velocity. This aeroelastic equation (\ref{eq:eq00}) is often transformed into modal coordinates to reduce the size of the computational problem and the unsteady generalized forces are approximated by rational functions in the Laplace domain \cite{CB14,CB15}. Using space-state formulation, or inverse Fourier transform,  the loads and structural forces (shear forces, bending moments, torques) are computed from internal loads given by \\

\begin{equation}
\textbf{F}(t)=-\textbf{M}\ddot{z}(t)-\textbf{F}_{a}(t)+\textbf{F}_{g}(t),
\label{eq:eq01}
\end{equation}

where $\textbf{F}_{a}(t)$ are the temporal unsteady aerodynamic forces and $\textbf{F}_{g}(t)$ are the temporal gust forces.\\

From the simulation model (\ref{eq:eq01}), the maximum temporal bending moment due to the discrete gust is computed for each flight point within the flight envelope. Airbus describes preliminary parameters by a vector of parameters $\textbf{x} \in \mathbb{R}^{d}$, with $d=20$. These preliminary parameters are: the aircraft mass, the zero fuel aircraft mass, the quantity of fuel, the true air speed, the Mach number,  the altitude, the load factors (NX, NY, NZ), the coordinates of the center of gravity (CGx, CGy, CGz) and the moments of inertia (Ixx, Ixy, Ixz, Iyy, Iyz, Izz), the wavelength ($H$) of the discrete gust and the flight profile alleviation factor $F_{g}$. Note that the the time history of the discrete gust has the following form \cite{CB10,CB11}

\begin{equation}
u(t)=\frac{U_{max}}{2}(1-\cos{\frac{2\pi t}{H}}),
\label{eq:eq1}
\end{equation}

with $0\leq t\leq \frac{H}{V}$, $H$ being the wavelength of the gust, $U_{max}$ is the maximum gust velocity given for a reference gust velocity $U_{ref}$ (depending on the altitude of the aircraft). $U_{max}$ satisfies 
\begin{equation}
U_{max}=U_{ref}\textit{F}_{g}(\frac{H}{350})^{\frac{1}{6}}.
\label{eq:eq2}
\end{equation}
$\textit{F}_{g}$ depends on the altitude, the maximum take off weight, the maximum landing weight and the maximum zero fuel weight. For each preliminary parameter $\textbf{x}$ and any location $k,\ 1\leq k \leq K$ ($K=45$ in our application, see Section 4) of the wing span, the maximum temporal bending moments $M_{B}(k,\textbf{x})$ constitute the initial database, i.e

\begin{equation}
M_{B}(k,\textbf{x})=\max_{0\leq t\leq \frac{2H}{V}} M(k,\textbf{x},t).
\label{eq:eqmax}
\end{equation}

For structural sizing, the quantities of interest are $M_{B}^{*}(k,\textbf{x})=\max_{\textbf{x}}M_{B}(k,\textbf{x})$ for all $k$, which is the maximum of the maximum temporal bending moment over the preliminary parameters envelope. It should be noticed that for a given aircraft, the critical preliminary parameter which gives the maximum of the maximum temporal bending moment can differ from a wing span location to an other. Moreover for a weight aircraft variant and for a given location on the wing span, the critical preliminary parameter is not necessarily the same than the one of the initial aircraft.\\

\section{Maximal Bending Moments vs Preliminary Parameter Response Surface}

Although there exist a lot of surrogate models such as GP-Kriging (see \cite{CB1,CB3}), Regression Trees \cite{fou18} or Neural Networks \cite{CB3} to approximate bending moments, we have chosen to develop a second order polynomial expansion. This response surface has the advantage to give interpretable results of the influence of the preliminary parameters. Moreover, for real world applications as we will see in the next section, it provides good results in terms of accuracy and sparseness. For a given location $k$ along the wing span and a preliminary parameter $\textbf{x}$, the maximal temporal bending moment is approximated by 
\begin{equation}
\hat{M}(k,\textbf{x})=a_{0}(k)+\sum_{i=1}^{d}a_{i}(k)x_{i}+ \sum_{i=1}^{d}\sum_{j=i}^{d}a_{i,j}(k)x_{i}x_{j}=\sum_{i=1}^{D}\omega_{i}(k)\Phi_{i}(\textbf{x}).
\label{eq:eq4}
\end{equation}

For each location $k$, the $\omega_{i}(k)$ are the regression coefficients to be estimated from the values of the initial database $M_{B}(k,\textbf{x})$ over the preliminary parameters envelope. The number of points of the preliminary parameters envelope is denoted by $n$ ($n=1560$ in our real world case study, see next section). The number of regression coefficients to be identified is $D=\frac{d^{2}+3d+2}{2}=231$ at each location $k$. This number of regression coefficients is quite low compared to the number $n$ of bending moments contained in the initial database. Instead of using Ordinary Least Square \cite{rencher2008} to estimate all the $D$ coefficients, we prefer to use an algorithm which is based on parsimony principles that aims to avoid overfitting. Among the existing algorithms such as Ridge \cite{marquardt75,tibshirani94}, Lasso \cite{tibshirani94}, we choose the Orthogonal Greedy Algorithm (OGA) \cite{barron2008, sancetta2016, mallat93} which makes a good balance between computer time due to its sequential structure and sparse representation due the direct control of the number of regression coefficients. \\

The Orthogonal Greedy Algorithm works as follows. From the initial database and for a given location $k$, $1\leq k \leq K$, along the wing span, we have $n$ maximum temporal bending moments $M_{B}(k,\textbf{x}_{i}), i=1,...,n$.  We build the matrix of preliminary parameters $\textbf{X}=[\textbf{x}_{1},...,\textbf{x}_{n}]^{\textbf{T}} \in \mathcal{M}_{n\text{x} d}(\mathbb{R})$ (a row of $\textbf{X}$ is a $\textbf{x}_{i}$) and the corresponding vector of maximum temporal bending moment at the station $k$ denoted by $\textbf{M}=(M_{B}(k,\textbf{x}_{1}),...,M_{B}(k,\textbf{x}_{n}))^{\textbf{T}}=(M_{1},...,M_{n})^{\textbf{T}}\in \mathbb{R}^{n}$. Knowing the different polynomial basis functions to be used, we can then build the matrix $\Phi=(\Phi_{j}(\textbf{x}_{i}))_{\substack{i=1,...,n \\ j=1,...,D}} \in \mathcal{M}_{n\text{x} D}(\mathbb{R})$.\\

The number $l \in \mathbb{N}^{*}, 1\leq l\leq D$, is the number of regression coefficients to be estimated and is fixed by the user. The OGA algorithm can be described by the following steps\\

\begin{algorithm}[H]
  \caption{Algorithm OGA}\label{euclid}
  \begin{algorithmic}[1]
  \State $\textit{Set: } l\in \mathbb{N}^{*}$
  \State $\textit{Set: } h_{0}=0$
  \For {$j=1,...,l$}:
    \State $s(j):=\underset{1\leq k\leq D}{\mathrm{arg min}} |\frac{1}{n}\sum_{i=1}^{n}(M_{i}-h_{0})\Phi_{k}(\textbf{x}_{i})|$
    \State $P^{j}:= \textit{OLS operator on } span\{\Phi_{s(1)},...,\Phi_{s(j)}\}$
    \State $h_{j}:=P^{j}\textbf{M}$
  \EndFor
  \end{algorithmic}
\end{algorithm}

$h_{j}$ is the projection of $M$  onto the orthogonal $span \{\Phi_{s(1)},...,\Phi_{s(j)}\}$. The aim of the algorithm is to choose the best $l$ functions among the $D$ initial basis functions which minimize the residuals. A Cross-Validation \cite{stone1974} technique is applied to reduce the number of unsignificant basis functions among the $l$. Notice that criteria such as AIC \cite{akaike} or BIC \cite{schwarz1978} can be also implemented. Once the regression coefficients are obtained by the OGA, the response surfaces are used to predict quickly the expected maximum temporal bending moment $\hat{M}(k,\textbf{x}_{wv})$ for each preliminary parameters $\textbf{x}_{wv,\ i},\ i=1,...,n_{wv}$ of the new aircraft variant and at any location $k$ of the wing span. Notice that the preliminary parameters of the weight variant are different from those of the initial database aircraft. From these values, the maximum of the expected maximum temporal values over the preliminary parameters envelop of the weight variant is directly extracted at each location $\hat{M}_{wv}^{*}(k)=\max_{i=1,...,n_{wv}}\hat{M}(k,\textbf{x}_{wv,\ i})$. Moreover, under normality assumption, the prediction intervals \cite{rencher2008} can be straightforwardly computed. All of these quantities of interest are then used for structural design studies.\\

\section{Numerical experiments}

The initial database comes from an Airbus aircraft having 235t as maximum take off weight with $K=45$ locations distributed along the wing span. This database is composed of $n=1560$ preliminary parameters representing the flight and structural envelope. Recall that each preliminary parameter is represented by 20 values. From each of these preliminary parameters, the temporal maximal bending values are computed along the 45 wing span locations. Using the OGA algorithm with $l=80$ and a Cross-Validation technique (6-folds), for each wing span location we have identified around 50 regression coefficients (instead of 231 coefficients when applying Ordinary Least Squares). The 210t and 280t are weight aircraft variants already developed and have the same wing geometry as the 235t. For each new weight variant, each database is composed of $n_{wv_{210}}=n_{wv_{280}}=1560$ preliminary parameters, but the preliminary parameters envelops are different. Therefore, it is possible to compare and validate the response surface predictions with reliable bending moments contained in the existing database of the two aircraft variants.\\

\begin{figure}[h!]
\centering
\subfigure[]{\includegraphics[width=10cm]{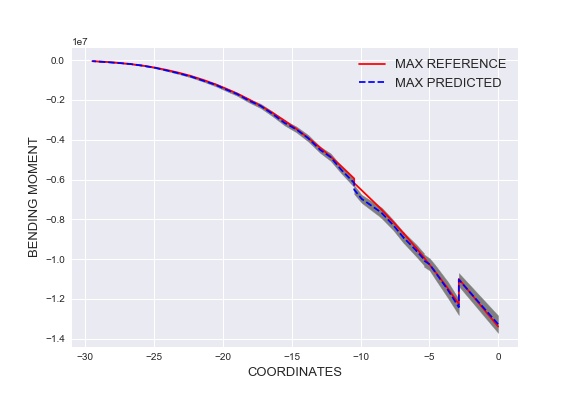}}\\
 \subfigure[]{\includegraphics[width=10cm]{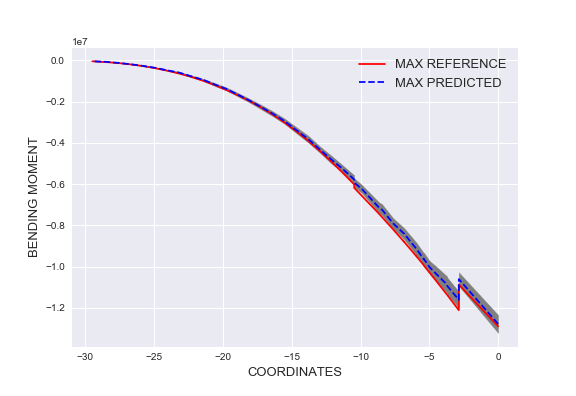}}\\
\caption{Maximum predicted bending moment envelope - the red line corresponds to the true maximum bending moment, the blue dashed line corresponds to the prediction, the grey zone corresponds to the prediction interval : (a) Extrapolation of 210t maximum, (b) Extrapolation of 280t maximum}
\label{label-Max}
\end{figure}

At any location $k$ of the wing span, using the associated surface responses we have computed the expected maximum of the maximum temporal bending moment of the new aircraft variants, denoted by $\hat{M}_{wv}^{*}(k)=\max_{i=1,...,n_{wv}} \hat{M}(k,\textbf{x}_{wv,\ i})$, over the preliminary parameter envelope (in blue in Figure \ref{label-Max}). This maximum $\hat{M}_{wv}^{*}(k)$ has been compared to the maximum bending moment $M_{B,\ wv}^{*}(k)$ given by the database (in red in Figure \ref{label-Max}).  The 210t aircraft gives a maximal relative error along the wing span of 2\% and a relative error at the wing root of 0.8\% when comparing the response surfaces predictions with the 210t existing database. The 280t aircraft gives a maximal relative error along the wing span of 3.5\% and a relative error at the wing root of 0.9\%. After having checked the normality of the residuals using Kolmogorov-Smirnoff and Shapiro-Wilk tests, we have computed the 99\% prediction interval bounds (see \cite{rencher2008}) provided by the response surfaces (in grey in Figure \ref{label-Max} (a-b)). It should be noted that the maximum bending moments provided by the two databases are quite close to the center of the prediction interval for any location of the wing span. Indeed, the maximal width of these prediction intervals is 8\% over the wing span. The above results obtained on real world aircraft cases show in particular that the second order response surfaces are quite competitive for preliminary design studies.\\

Besides the good prediction results in terms of preliminary design study, the proposed methodology requires a short computational time on a standard desk computer. It has taken around 10 minutes for computing the expected and the prediction intervals of the maximum bending moments along the wing span for the two weight variants. More precisely, we have built $45$ surface responses (i.e the coefficients for each location) from the $1560\times45=70200$ initial data of the 235t. We have predicted the $1560\times45\times2=140400$ temporal maximum bending moments of the 210t and 280t. Finally, we have extracted from the response surface both the expected maximum of the maximum temporal bending moment and the associated prediction intervals at each location of the wing span for the two aircraft variants.

\section{Conclusion}

This note has presented a reliable and fast methodology for estimating critical load cases for weight variants aircraft at the preliminary design phase. After having identified the set of preliminary parameters, the proposed methodology used the Orthogonal Greedy Algorithm to identify the coefficient of the second order polynomial response surfaces from an initial database aircraft from which the weight variants aircraft belongs. This Orthogonal Greedy Algorithm is highly efficient in terms of computational time and parsimony representation for estimating the regression coefficients. We reported very encouraging results concerning a real world case of wing bending moments. Indeed, having identified response surfaces for the 235t  Airbus aircraft, the maximal errors predictions are 2\% for the 210t variant and 3.5\% for the 280t variant.
Future related work will attempt to extend our approach to other structural parts and flight conditions (i.e continuous turbulence, landings, fuselage, ...).

\bibliographystyle{spmpsci}  

\bibliography{biblio}

\begin{thebibliography}{10}
\providecommand{\url}[1]{{#1}}
\providecommand{\urlprefix}{URL }
\expandafter\ifx\csname urlstyle\endcsname\relax
  \providecommand{\doi}[1]{DOI~\discretionary{}{}{}#1}\else
  \providecommand{\doi}{DOI~\discretionary{}{}{}\begingroup
  \urlstyle{rm}\Url}\fi

\bibitem{akaike}
Akaike, H.: Information theory and an extension of the maximum likelihood
  principle.
\newblock Second International Symposium on Information Theory pp. 267--281
  (1973)

\bibitem{barron2008}
Barron, A.R., Cohen, A., Dahmen, W., DeVore, R.A.: Approximation and learning
  by greedy algorithms.
\newblock Ann. Statist. \textbf{36}(1), 64--94 (2008)

\bibitem{CB6}
Castellani, M., Cooper, J.E., Lemmens, Y.: Flight loads prediction of high
  aspect ratio wing aircraft using multibody dynamics.
\newblock International Journal of Aerospace Engineering \textbf{2016} (2016)

\bibitem{CB7}
Castrichini, A., Cooper, J.E., Benoit, T., Lemmens, Y.: Gust and ground loads
  integration for aircraft landing loads prediction.
\newblock In: 58th AIAA/ASCE/AHS/ASC Structures, Structural Dynamics, and
  Materials Conference, p. 0636 (2017)

\bibitem{CB14}
Cotoi, I., Botez, R.M.: Method of unsteady aerodynamic forces approximation for
  aeroservoelastic interactions.
\newblock Journal of guidance, control, and dynamics \textbf{25}(5), 985--987
  (2002)

\bibitem{CB2}
Dharmasaroja, A., Armstrong, C., Murphy, A., Robinson, T., McGuinness, S.,
  Iorga, N., Barron, J.: Load case characterization for the aircraft structural
  design process.
\newblock AIAA Journal pp. 2783--2792 (2017)

\bibitem{CB10}
Duven, J.E.: Advisory circular - dynamic gust loads (2014)

\bibitem{fou18}
Fournier, E., Grihon, S., Klein, T.: {A case study : Influence of Dimension
  Reduction on regression trees-based Algorithms -Predicting Aeronautics Loads
  of a Derivative Aircraft} (2018).
\newblock \urlprefix\url{https://hal.archives-ouvertes.fr/hal-01700314}.
\newblock Working paper or preprint

\bibitem{CB1}
Haddad~Khodaparast, H., Georgiou, G., Cooper, J., Riccobene, L., Ricci, S.,
  Vio, G., Denner, P.: Efficient worst case "1-cosine" gust loads prediction.
\newblock Journal of Aeroelasticity and Structural Dynamics \textbf{2}(3)
  (2012)

\bibitem{howe2004}
Howe, D.: Aircraft loading and structural layout.
\newblock AIAA Education Series. (2004)

\bibitem{CB13}
Karpel, M., Strul, E.: Minimum-state unsteady aerodynamic approximations with
  flexible constraints.
\newblock Journal of Aircraft \textbf{33}(6), 1190--1196 (1996)

\bibitem{CB4}
Khodaparast, H.H., Cooper, J.: Rapid prediction of worst-case gust loads
  following structural modification.
\newblock AIAA journal \textbf{52}(2), 242--254 (2014)

\bibitem{CB5}
Knoblach, A., Looye, G.: Efficient determination of worst-case gust loads using
  system norms.
\newblock Journal of Aircraft \textbf{54}(3), 1205--1210 (2017)

\bibitem{lomax1996}
Lomax, T.L.: Structural loads analysis for commercial transport aircraft:
  theory and practice.
\newblock American Institute of Aeronautics and Astronautics (1996)

\bibitem{mallat93}
Mallat, S.G., Zhang, Z.: Matching pursuits with time-frequency dictionaries.
\newblock IEEE Transactions on Signal Processing \textbf{41}(12), 3397--3415
  (1993)

\bibitem{marquardt75}
Marquardt, D.W., Snee, R.D.: Ridge regression in practice.
\newblock The American Statistician \textbf{29}(1), 3--20 (1975).
\newblock \urlprefix\url{http://www.jstor.org/stable/2683673}

\bibitem{CB3}
Nazzeri, R., Haupt, M., Lange, F., Sebastien, C.: Selection of Critical Load
  Cases Using an Artificial Neural Network Approach for Reserve Factor
  Estimation.
\newblock Deutsche Gesellschaft fur Luft-und Raumfahrt-Lilienthal-Oberth eV
  (2015)

\bibitem{papila2000}
Papila, M., Haftka, R.T.: Response surface approximations: noise, error repair,
  and modeling errors.
\newblock AIAA journal \textbf{38}(12), 2336--2343 (2000)

\bibitem{CB15}
Poirion, F.: Multi-mach rational approximation of generalized aerodynamic
  forces.
\newblock Journal of aircraft \textbf{33}(6), 1199--1201 (1996)

\bibitem{rencher2008}
Rencher, A.C., Schaalje, G.B.: Linear models in statistics.
\newblock John Wiley \& Sons (2008)

\bibitem{sancetta2016}
Sancetta, A.: Greedy algorithms for prediction.
\newblock Bernoulli \textbf{22}(2), 1227--1277 (2016)

\bibitem{schwarz1978}
Schwarz, G.: Estimating the dimension of a model.
\newblock The annals of statistics \textbf{6}(2), 461--464 (1978)

\bibitem{stone1974}
Stone, M.: Cross-validatory choice and assessment of statistical predictions.
\newblock Journal of the royal statistical society. Series B (Methodological)
  pp. 111--147 (1974)

\bibitem{tibshirani94}
Tibshirani, R.: Regression shrinkage and selection via the lasso.
\newblock Journal of the Royal Statistical Society, Series B \textbf{58},
  267--288 (1994)

\bibitem{CB11}
Zhao, Y., Yue, C., Hu, H.: Gust load alleviation on a large transport airplane.
\newblock Journal of Aircraft \textbf{53}(6), 1932--1946 (2016)

\end{thebibliography}

\end{document}